\documentclass{PoS}
\usepackage[]{latexsym}
\usepackage[]{amsfonts}
\usepackage[]{amsmath}
\usepackage[]{graphicx}
\usepackage[]{epsf}

\newcommand{\Tr}{\textrm{Tr}}

\newcommand{\ev}[1]{\langle #1 \rangle}

\newcommand{\dd}{\textrm{d}}

\newcommand{\SE}{S_{\textrm{E}}}
\newcommand{\MSb}{\overline{\textrm{MS}}}

\newcommand{\mub}{\bar\mu}
\newcommand{\Nc}{N_{\textrm{c}}}
\newcommand{\Nf}{N_{\textrm{f}}}

\title{Quark number susceptibility of high temperature and finite density QCD.}

\ShortTitle{Quark number susceptibility}

\author{\speaker{A. Hietanen}\\
  Theoretical Physics Division, Department of Physical Sciences P.O. Box 64 FI-00014 University
  of Helsinki, Finland and \\
  Helsinki Institute of Physics, P.O. Box 64, FI-00014 University of Helsinki, Finland \\        E-mail: \email{ari.hietanen@helsinki.fi}}

\author{K. Rummukainen\\
  Department of Physics, University of Oulu P.O. Box 3000, FI-90014 Oulu \\
  E-mail: \email{kari.rummukainen@oulu.fi}}

\abstract{We utilize lattice simulations of 
the dimensionally 
reduced effective field theory (EQCD)
to determine the quark number susceptibility of QCD at high temperature
($T>2T_c$).  We also use analytic continuation to obtain results at
finite density.  The results extrapolate well from known perturbative
expansion (accurate in extremely high temperatures) to 4d lower
temperature lattice data.}

\FullConference{The XXV International Symposium on Lattice Field Theory\\
		 July 30-4 August 2007\\
		 Regensburg, Germany}

\begin{document}

\section{Introduction}
\begin{figure}[b]
  \begin{center}
  \includegraphics*[width=\textwidth]{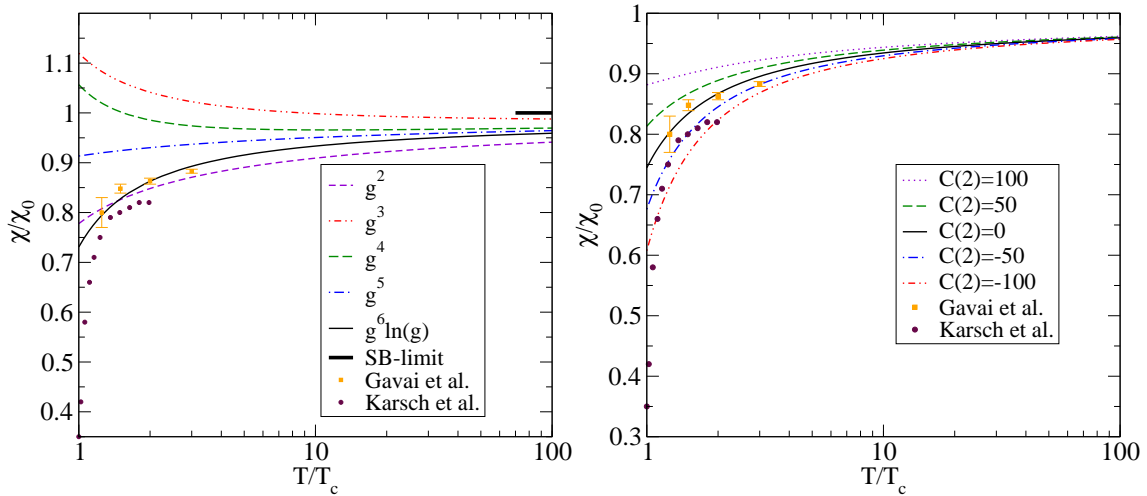}
  \caption{{\em Left:} The perturbative expansion of quark number
    susceptibility order by order for $N_f=2$. The coefficient
    at order $g^6$ has been fixed here to match the lattice
    measurements. 
    {\em Right:} 
    The effect of changing the 
    value of the unknown $O(g^6)$ coefficient, parametrised as
    $C(N_f)\left(\frac{g^2}{4\pi^2}\right)^3$. The perturbative
    results are from \cite{vuorinen02} and the lattice results from \cite{gavai,karsch}.}
\label{soc}
\end{center}
\end{figure}

The quark number (baryon number) susceptibility is an observable which
is one of the signature of the quark-gluon plasma in heavy ion
collision experiments \cite{Asakawa:2000wh}.  Thus, it is of interest to
accurately calculate the quark number susceptibility theoretically.
In weak-coupling perturbation theory, the susceptibility of the
quark-gluon plasma has been calculated up to order $g^6\ln 1/g$ \cite{vuorinen02}.  Because of the asymptotic freedom, at high
enough temperatures the perturbation theory is a valid approach.
However, the convergence of the perturbative series is bad at
physically accessible temperatures, and the applicability of the
results is not obvious, see Fig.~\ref{soc}.  The order-by-order
behaviour of the susceptibility is not systematic, and the
low-temperature behaviour changes qualitatively at orders $g^3$ and
$g^5$.  Further, if we allow variations in the unknown
$O(g^6)$-coefficient in the expansion (Fig.~\ref{soc} right), we obtain
quite large variation in the final result at temperatures under $10T_c$.

The quark number susceptibility has also been studied with lattice
simulations \cite{gavai,karsch}.  While the standard lattice methods
(with dynamical fermions) are the best method to study QCD in the
immediate vicinity of the phase transition, at higher temperatures it
is more economical to simulate dimensionally reduced effective theory,
electrostatic QCD (EQCD) \cite{klrs}.  This method has already been
used to calculate the pressure in quark-gluon plasma \cite{resum}.
Here we present the updated results of simulations of quark number
susceptibility for $N_f=2$ \cite{suskis06}
for zero chemical potential.  We also present preliminary results from
simulations extended to finite chemical potential. These results are
obtained by doing simulations with imaginary values of the chemical
potential and then analytically continuing to the real values.  This
is achieved by fitting a polynomial of $\mu^2$ to the data. 

\section{Susceptibility in electrostatic QCD}
EQCD is defined by the action
\begin{eqnarray}
       \SE & = & \int \dd^3x\left\{
          \frac{1}{2}\Tr[F_{ij}^2]+\Tr[D_i,A_0]^2+m_3^2\Tr[A_0^2]+
      i\gamma_3 \Tr[A_0^3] + \lambda_3(\Tr[A_0^2])^2 \right\},
      \label{action}
\end{eqnarray}
where $F_{ij}=\partial_i A_j - \partial_j A_i + ig_3[A_i,A_j]$ and
$D_i=\partial_i+ig_3A_i$. $F_{ij}$, $A_i$ and $A_0$ are traceless
$3\times 3$ Hermitean matrices ($A_0=A_0^aT_a$, etc).  The coupling
and the mass parameters $g_3$, $m_3$, $\gamma_3$ and $\lambda_3$ are
determined by the physical 4d temperature, renormalization scale
$\Lambda_{\MSb}$, chemical potential $\mu$ and the number of massless
fermions.  It is convenient to use the dimensionless ratios
\begin{equation}
  y  =  \frac{m_3^2}{g_3^4}, ~~~~~ x = \frac{\lambda_3}{g_3^2},
  ~~~~~ z=\frac{\gamma_3}{g_3^3},
\end{equation}
which determine the physical properties of EQCD.  The 
$\mu$-dependence of the parameters is, at 1-loop level,
\begin{equation}
         y = y_{\mu=0} \left(1+\sum_f\mub_f^2\frac{3}{2\Nc+\Nf}\right),
  ~~~~~~ z = \sum_f \frac{\mub_f}{3\pi},
  ~~~~~~ x = x_{\mu=0}\,,
\end{equation}
where $\mub = \mu/(\pi T)$ and the $\mu=0$ expressions can 
be found in ref.~\cite{klrs}.
The two loop corrections have been calculated in ref.~\cite{hart00},
but the effects remain in practice negligible.

The quantity we are interested in is the quark number susceptibility,
which we define as a derivative over one flavor $u$ only:
\begin{equation}
  \chi_3  \equiv  \frac{1}{V}\frac{\partial^2}{\partial \mu_u^2}\ln{\cal Z} 
   =   \frac{1}{V}\frac{\partial^2}{\partial \mu_u
    ^2}\ln\int \mathcal{D}A_kA_0\exp\left(-\SE\right)
\end{equation}
Substituting $\SE$ from (\ref{action}) we arrive at the result
\begin{eqnarray}
  \chi_3& = & -\frac{6}{2N_c+N_f}\,y_{\mu=0}\, \ev{\Tr A_0^2} 
  \nonumber \\ & & 
  + \frac{1}{V 9\pi^2} \int d^3r_1 d^3r_2  
   \left(\ev{\Tr A_0^3(r_1)\Tr A_0^3(r_2)}-\ev{\Tr A_0^3}^2 \right) 
   \nonumber \\ & & 
  + \frac{36}{V(2N_c+N_f)^2}\mub_u^2\,y_{\mu=0}^2\,
  \int d^3 r_1 d^3r_2 \left(\ev{\Tr A_0^2(r_1)\Tr A_0^2(r_2)} - 
    \ev{\Tr A_0^2}^2\right).
  \label{suskis}
\end{eqnarray}
Thus, the quark number susceptibility is obtained by measuring the
condensates $\ev{Tr A_0^2}$, $\ev{(\Tr A_0^2)^2}$ and 
$\ev{(\Tr A_0^3)^2}$ on the lattice.  Due to the
superrenormalizable nature of the theory,  measurements can
be rigorously converted to $\MSb$ scheme in the lattice continuum limit,
and because $\MSb$ was used in the perturbative matching to 4d QCD,
this also allows us to compare to 4d results.

The lattice counterterms needed for the continuum limit of $\ev{Tr A_0^2}$ are
given in \cite{klrs}, and of $ V\ev{\Tr A_0^3(r_1)\Tr A_0^3(r_2)}$ in
\cite{suskis06}.  The contribution including 
$\ev{\Tr A_0^2(r_1)\Tr A_0^2(r_2)}$ is not UV divergent and thus
does not require counterterms.

Finally, the relation between $\chi_3$ and the true 4d susceptibility is
\begin{equation}
  \chi = \frac{g_3^6}{T^3}\chi_3 + \frac{\partial^2}{\partial
  \mu_u^2}\Delta p,
  \label{eqsusc}
\end{equation}
where $\Delta p=p_\textrm{QCD}-p_\textrm{3d}$ is the perturbative
3d$\rightarrow$4d matching coefficient for pressure, and can
be found in~\cite{vuorinen03}. 

\section{Lattice measurements}
Lattice simulations are carried out for $\Nf=2$. We use 6
different values of chemical potential $\mu$, 
and for each value of $\mu$ we use
eight different values of temperature $T$.  For 
each of these $(\mu,T)$-pairs we use five different values of 
the lattice spacing $a$, in order to obtain
a reliable continuum limit.  To check the finite volume effects we did
simulations with different volumes at the smallest lattice spacing; for
a detailed analysis in a related theory see \cite{plaquette}.

\begin{figure}
  \begin{center}
  \includegraphics*[width=0.4\textwidth]{a2sdclz0.eps}
  \includegraphics*[width=0.4\textwidth]{a2sdclz01.eps}
  \includegraphics*[width=0.4\textwidth]{A3sdclz0.eps}
  \includegraphics*[width=0.4\textwidth]{A3sdclz01.eps}
  \caption{Continuum extrapolations of $V(\ev{(\Tr
      A_0^{2})^2}-\ev{\Tr A_0^2}^2)$ and $V(\ev{(\Tr
      A_0^3)^2}-\ev{\Tr A_0^3}^2)$.}
\label{clsd}
\end{center}
\end{figure}

\begin{figure}
  \begin{center}
    \includegraphics*[width=0.65\textwidth]{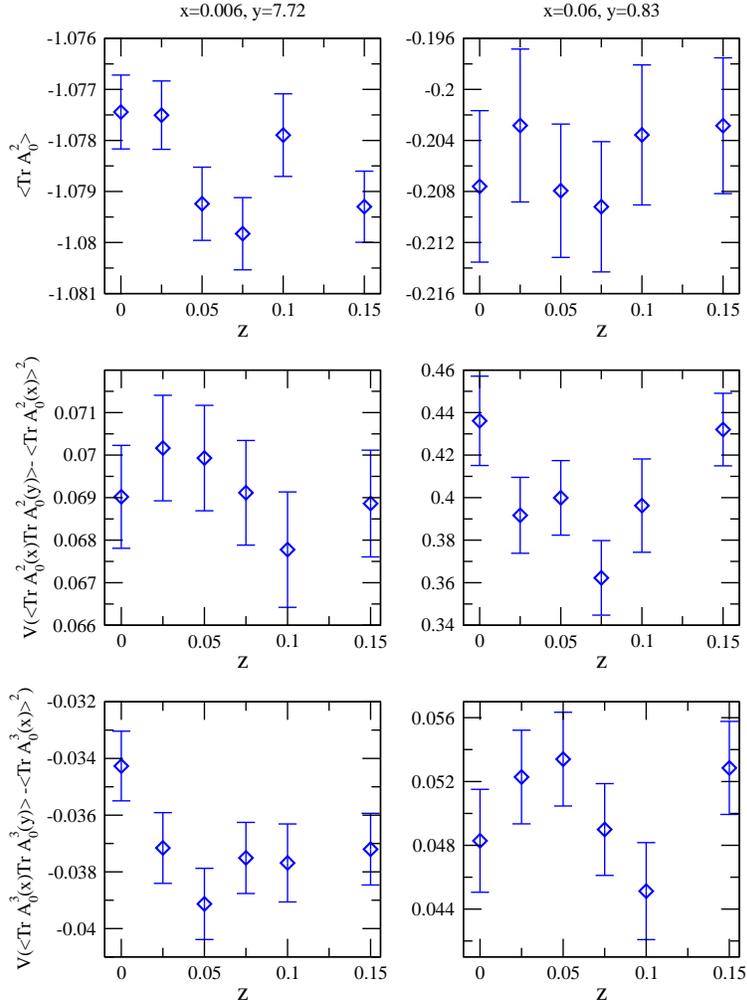}
    \caption{Simulations done with two different $(x,y)$-pairs
      while varying the imaginary chemical potential in the
      range $0\leq -iz \leq 0.15$. 
      Within our accuracy varying $iz$ does change the results,
      making analytic continuation straightforward.}
   \end{center}
  \label{zdiff}
\end{figure}

\begin{figure}[p]
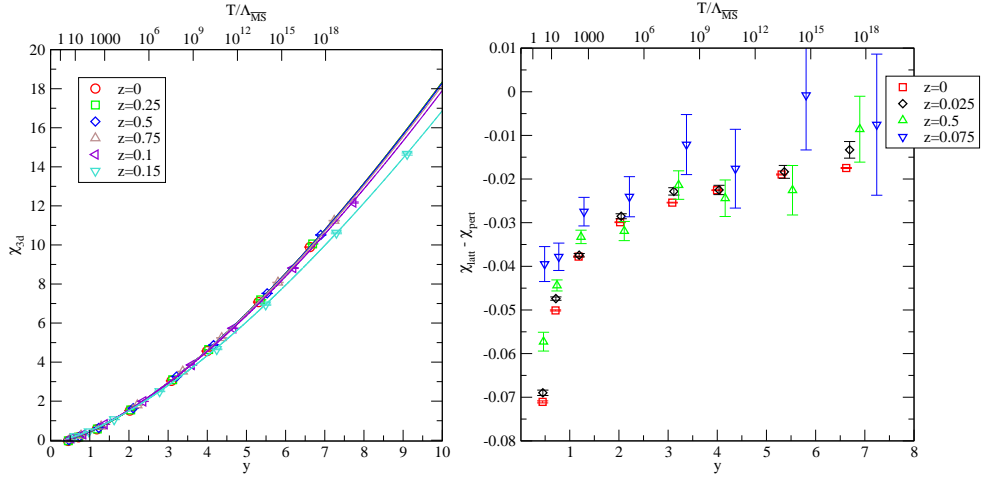

  \begin{center}
    \includegraphics*[width=0.39\textwidth]{suskis3d.eps}
    \includegraphics*[width=0.46\textwidth]{suskis3dres.eps}
    \caption{Left: The quark number susceptibility in EQCD, with
      different values of chemical potential in dimensionless units. The solid
      lines are the perturbative result.  Right: The difference between
      the perturbative and lattice results.  (The
      $T$-scale on the top of the figures corresponds to $z=0$ case.)}
    \end{center}
  \label{suskis3d}
\end{figure}

\begin{figure}[p]
  \begin{center}
    \includegraphics*[width=0.7\textwidth]{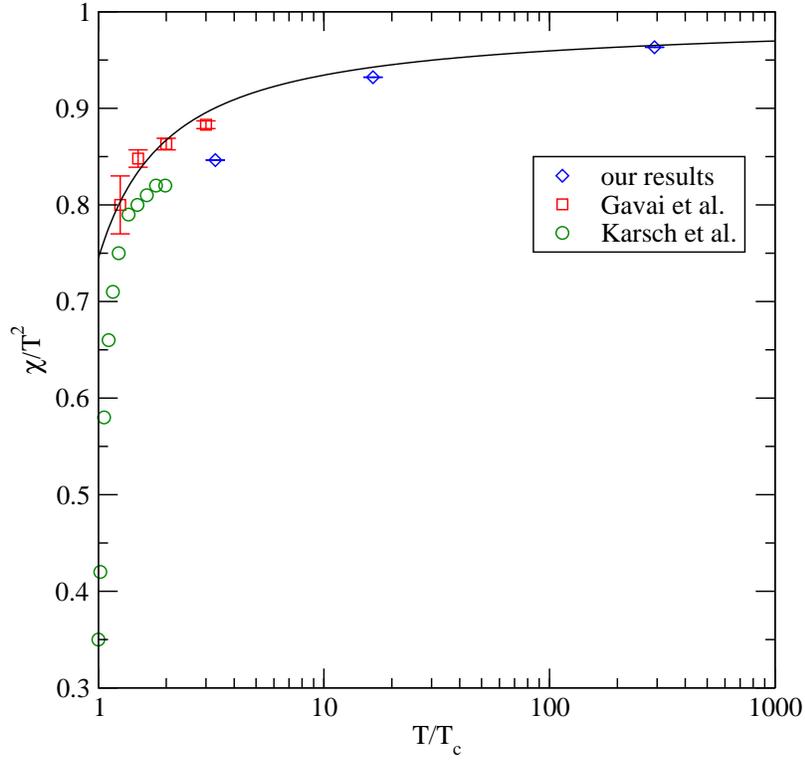}
    \caption{The susceptibility in 4d units at $\mu=0$. 
      The results  agree with 4d lattice simulation results.}
    \label{suskis4d}
  \end{center}
\end{figure}

Precise continuum limits are necessary for accurate determination
of $\chi_3$.  For the condensate $\ev{\Tr A_0^2}$
we use a fit ansatz of form
\begin{equation} 
  c_1+\frac{c_2}{\beta}+{\frac{c_2'}{\beta}\log(\beta)}+\frac{c_3}{\beta^2},
\label{logextra}
\end{equation}
where $\beta = 6/(g_3^2 a)$.
The existence of the logarithmic term increases the final errors
significantly. However, the coefficients of the above ansatz are
perturbatively calculable, and there is an ongoing program to determine
the coefficient of the log-term using stochastic perturbation theory
\cite{Torrero:2006sj}.  The knowledge
of this would reduce the errors by order of magnitude.

The contributions $V(\ev{(\Tr A_0^{2})^2}-\ev{\Tr A_0^2}^2)$ and
$V(\ev{(\Tr  A_0^3)^2}-\ev{\Tr A_0^3}^2)$ are fitted with second order
polynomial ansatz
\begin{equation} 
  c_1+\frac{c_2}{\beta}+\frac{c_3}{\beta^2}.
  \label{polyextra}
\end{equation}
This fits the data well, see Fig.~\ref{clsd}.

We study the chemical potential dependence by using imaginary $\mu$
and performing analytical continuation.  However, this turns out to be
rather trivial: for fixed $x,y$ the dependence of the results on $iz
\propto i\mu$ is very small and not visible within our statistical
errors, see Fig.~\ref{zdiff}.  (However, see the note above about the
statistical errors in $\ev{A_0^2}$.)
Thus, the $\mu$-dependence of the results is, in practice, completely 
due to the $\mu$-dependence of $y$.

The final continuum extrapolated results agree well with
the perturbative susceptibility.  It is of the form
\begin{equation}
  \chi_{\textrm{pert}}=a_1y^{3/2}+a_2y+a_3y^{1/2}+a_4.
\end{equation}
Hence the difference of lattice and perturbation theory should behave
as $y^{-1/2}$, which is the case, as can be seen in Fig.~\ref{suskis3d}.

After the matching to 4d QCD we obtain the susceptibility in physical
units. Our results significantly deviate downwards from the
perturbation theory, as can be seen from Fig.~\ref{suskis4d}, bringing
the results closer to the recent simulations by Karsch et
al.~\cite{karsch}.  However, one should bear in mind that our results
suffer from a matching ambiguity related to the unknown $O(g^6)$
coefficient in perturbation theory, see Fig.~\ref{soc}, and the
results need to be matched to a known point (4d lattice simulation) at
some low temperature.
Nevertheless, we can say that the deviation from perturbative
result is still rather large at $10T_c$. See Fig.\ref{suskis4d}.

\acknowledgments

This work has been partly supported by the Magnus Ehrnrooth
Foundation, a Marie Curie Felloswhip for Early Stage Researchers
Training, and the Academy of Finland, contract number 104382.
Simulations have been carried out at the Finnish IT Center for Science
(CSC).

\end{document}